\documentclass[12pt,aps,groupedaddress,superscriptaddress]{revtex4}
\usepackage{graphicx}
\usepackage{amsmath,amssymb,amsfonts}
\usepackage{natbib}

\newcommand{\be}{\begin{equation}}
\newcommand{\ee}{\end{equation}}

\begin{document}

\title{Time-Dependent Density Functional Theory \\
from a Bohmian Perspective}

\author{\'Angel S. Sanz}

\affiliation{Instituto de F\'{\i}sica Fundamental,
Consejo Superior de Investigaciones Cient\'{\i}ficas,
Serrano 123, 28006 Madrid, Spain}

\author{Xavier Gim\'enez}

\affiliation{Departament de Qu\'\i mica F\'\i sica,
Universitat de Barcelona i Parc Cientific de Barcelona,
Mart\'\i\ i Franqu\`es 1, 08028 Barcelona, Spain}

\author{Josep Maria Bofill}

\affiliation{Departament de Qu\'\i mica Org\'anica,
Universitat de Barcelona i Parc Cientific de Barcelona,
Mart\'\i\ i Franqu\`es 1, 08028 Barcelona, Spain}

\author{Salvador Miret-Art\'es}

\affiliation{Instituto de F\'{\i}sica Fundamental,
Consejo Superior de Investigaciones Cient\'{\i}ficas,
Serrano 123, 28006 Madrid, Spain}

\date{\today}

\begin{abstract}
This article has been published as a chapter in ``Chemical Reactivity
Theory: A Density Functional View'', ed. P. K. Chattaraj (CRC Press,
New York, 2009), ch. 8, p. 105.
In it, an overview of the relationship between time-dependent DFT and
quantum hydrodynamics is presented, showing the role that Bohmian
mechanics can play within the ab-initio methodology as both a numerical
and an interpretative tool.
\end{abstract}


\maketitle



\section{Introduction}
\label{sec1}

Since the early days of Quantum Mechanics, the wave-function theory has
proven to be very successful in describing many different quantum
processes and phenomena.
However, in many problems of Quantum Chemistry and
Solid State Physics, where the dimensionality of the systems studied is
relatively high, {\it ab initio} calculations of the structure of
atoms, molecules, clusters and crystals, and their interactions
are very often prohibitive.
Hence, alternative formulations based on the direct use of the
probability density, gathered under what is generally known as the
density matrix theory \cite{Blum}, were also developed since the very
beginning of the new mechanics.
The independent electron approximation or Thomas-Fermi model, and the
Hartree and Hartree-Fock approaches are former statistical models
developed in that direction \cite{szabo}.
These models can be considered direct predecessors of the more recent
density functional theory (DFT) \cite{koch}, whose principles were
established by Hohenberg, Kohn and Sham \cite{Kohn1,Kohn2} in the
middle sixties.
According to this theory, the fundamental physical information about
a many-body system is provided by single-particle densities in a
three-dimensional space, which are obtained variationally within a
time-independent framework.
When compared with other previous formalisms, DFT presents two clear
advantages: (i) it is able to treat many-body problems in a
sufficiently accurate way and (ii) it is computationally simple.
This explains why it is one of the most widely used theories to deal
with electronic structure ---the electronic ground-state energy as a
function of the position of the atomic nuclei determines the structure
of molecules and solids, providing at the same time the forces acting
on the atomic nuclei when they are not at their equilibrium positions.
At present, DFT is being used routinely to solve many problems in gas
phase and condensed matter.
Furthermore, it has made possible the development of accurate molecular
dynamics schemes in which the forces are evaluated quantum-mechanically
``on the fly''.
Nonetheless, DFT is a fundamental tool provided the systems studied are
relatively large; for small systems standard methods based on the use
of the wave function render quite accurate results \cite{Mcweeny}.
Moreover, it is also worth stressing that all practical applications
of DFT rely on essentially uncontrolled approximations \cite{Improv}
(e.g., the local density approximation \cite{Kohn1,Kohn2}, the local
spin-density approximation or generalized gradient approximations
\cite{GGA}), and therefore the validity of DFT is conditioned to its
ability to provide results sufficiently close to the experimental data.

As mentioned above, standard DFT is commonly applied to determine
ground states in time-independent problems.
Hence, reactive  and non-reactive scatterings as well as atoms and
molecules in laser fields have been out of the reach of the
corresponding  methodology.
Nevertheless, though it is less known than the standard DFT, very
interesting work in this direction can also be found in the literature
\cite{bloch,Bartolotti,Deb,Runge,DebChattaraj,Mcclendon,march}, where
DFT is combined with quantum hydrodynamics [or quantum-fluid dynamics
(QFD)] (QFD-DFT) in order to obtain a quantum theory of many-electron
systems.
In this case, the many-electron wave function is replaced by
single-particle charge and current densities.
The formal grounds of QFD-DFT rely on a set of hydrodynamical equations
\cite{Bartolotti,Runge,Deb}.
It has the advantage of dealing with dynamical processes evolving in
time in terms of single-particle time-dependent (TD) equations, as
derived by different authors \cite{march}.
Apart from QFD-DFT, there are other TD-DFT approaches based on similar
grounds, such as the Floquet DFT \cite{maitra,samal} or the quantal
DFT \cite{Sahni}.
Furthermore, we would like to note that TD-DFT does not necessarily
require to pass through a QFD or QFD-like formulation in order to be
applied \cite{botti}.
As happens with standard DFT, TD-DFT can also be started directly from
the many-body TD Schr\"odinger equation, the density being then
determined from solving a set of TD Schr\"odinger equations for single,
non-interacting particles \cite{Runge}.

Although trajectories are not computed in QFD-DFT, it is clear that
there is a strong connection between this approach and the trajectory
or hydrodynamical picture of quantum mechanics \cite{Holland},
independently developed by Madelung \cite{Madelung}, de Broglie
\cite{Broglie} and Bohm \cite{Bohm}, and which is also known as
Bohmian mechanics.
From the same hydrodynamical equations, information not only about
the system configuration (DFT calculations) but also about its
dynamics (quantum trajectories) is possible to obtain.
This fact is better understood when the so-called quantum potential
is considered, since it allows us to associate the probability density
(calculated from DFT) with the quantum trajectories.
Note that this potential is determined by the curvature
of the probability density and, at the same time, it governs the
behavior displayed by the quantum trajectories.
Because of the interplay between probability density and quantum
potential, the latter conveys fundamental physical information: it
transmits the nonseparability contained in the probability density
(or, equivalently, the wave function) to the particle dynamics.
This property, on the other hand, is connected with the inherent
nonlocality of Quantum Mechanics \cite{Bell}, i. e., two distant
parts of an entangled or nonfactorizable system will keep a strong
correlation due to coherence exhibited by its quantum evolution.

The purpose of this chapter is to show and discuss the connection
between TD-DFT and Bohmian mechanics, as well as the sources of lack
of accuracy in DFT, in general, regarding the problem of correlations
within the Bohmian framework or, in other words, of entanglement.
In order to be self-contained, a brief account of how DFT tackles the
many-body problem with spin is given in Sec.~\ref{sec2}.
A short and simple introduction to TD-DFT and its quantum
hydrodynamical version (QFD-DFT) is presented in Sec.~\ref{sec3}.
The problem of the many-body wave function in Bohmian mechanics, as
well as the fundamental grounds of this theory, are described and
discussed in Sec.~\ref{sec4}. This chapter is conluded with a short
final discussion in Sec.~\ref{sec5}.


\section{The many-body problem in standard DFT}
\label{sec2}

There are many different physical and chemical systems of interest
which are characterized by a relatively large number of degrees of
freedom.
However, in most of cases, the many-body problem can be reduced to
calculations related to a sort of inhomogeneous gas, i.e., a set of
{\it interacting} point-like particles which evolve
quantum-mechanically under the action of a certain
effective potential field.
This is the typical DFT scenario, with an ensemble of $N$ electrons in
a nuclear or external potential representing the system of interest.
DFT thus tries to provide an alternative approach to the exact,
nonrelativistic $N$-electron wave function $\Psi({\bf r}_1s_1, \ldots,
{\bf r}_Ns_N)$, which satisfies the time-independent Schr\"odinger
equation and where ${\bf r}_N$ and $s_N$ are the space and spin
coordinates, respectively.
Because the methodology based on DFT is easy and computationally
efficient in its implementation, this theory is still enjoying an
ever-increasing popularity within the Physics and Chemistry
communities involved in many-body calculations.

To understand the main idea behind DFT, consider the following.
In the absence of magnetic fields, the many-electron Hamiltonian does
not act on the electronic spin coordinates, and the antisymmetry and
spin restrictions are directly imposed on the wave function
$\Psi({\bf r}_1s_1,\ldots,{\bf r}_Ns_N)$.
Within the Born-Oppenheimer approximation, the energy of an
$N$-electron system with a fixed $M$-nuclei geometry ${\bf R}$
takes the following form in atomic units:
\begin{equation}
 E = - \frac{1}{2} \int_{{\bf r}_1 = {\bf r}_1'}
 \left[ \nabla \cdot \nabla^T \gamma_1({\bf r}_1;{\bf r}_1') \right]
  \! d{\bf r}_1
 + \int v_{\rm ext} ({\bf R},{\bf r}_1) \ \! \gamma_1 ({\bf r}_1)
   \ \! d{\bf r}_1
 + \int \frac{\gamma_2({\bf r}_1,{\bf r}_2)}{r_{12}} \
   d{\bf r}_1 \ \! d{\bf r}_2 ,
 \label{eq1prb}
\end{equation}
where $\gamma_1({\bf r}_1)$ and $\gamma_2({\bf r}_1,{\bf r}_2)$
are the diagonal elements of $\gamma_1({\bf r}_1;{\bf r}_1')$ and
$\gamma_2({\bf r}_1,{\bf r}_2;{\bf r}_1',{\bf r}_2')$, respectively,
which represent the one-electron (or one-particle) density and the
electron-electron (or two-particle) correlation function, commonly
used in DFT and electronic structure theory.
In principle, it might seem that all the information about the system
necessary to evaluate the energy is contained in $\gamma_1({\bf r}_1)$
and $\gamma_2({\bf r}_1;{\bf r}_2)$, and therefore one could forget
about manipulating the wave function.
However, in order to avoid unphysical results in the evaluation of
the energy, it is still necessary to compute the wave function
$\Psi({\bf r}_1s_1,\ldots,{\bf r}_Ns_N)$ that generates the correct
$\gamma_1({\bf r}_1)$ and $\gamma_2({\bf r}_1;{\bf r}_2)$ densities.
Equation~(\ref{eq1prb}) is the starting point of DFT, which aims to
replace both $\gamma_1({\bf r}_1;{\bf r}_1')$ and
$\gamma_2({\bf r}_1,{\bf r}_2)$ by $\rho ({\bf r})$.
If we are only interested in the system ground state, the
Hohenberg-Kohn theorems state that the exact ground-state total
energy of any many-electron system is given by a universal, unknown
functional of the one-electron density.
However, only the second term of Eq.~(\ref{eq1prb}) is an explicit
functional of $\rho ({\bf r})$.
The first term corresponds to the kinetic energy, which is a functional
of the complete one-electron density function
$\gamma_1({\bf r}_1;{\bf r}_1')$.
For $N$-electron systems the most important contribution to the
electron-electron term comes from the classical electrostatic
self-energy of the charge interaction, which is an explicit functional
of the diagonal one-electron function.
The remaining contribution to the electron-electron term is still
unknown.
These two terms are a functional of the one-electron density, namely
the ``exchange-correlation'' functional.
Thus, it is possible to define a universal functional which is
derivable from the one-electron density itself and with no reference
to the external potential $v_{ext}({\bf R},{\bf r})$.
According to McWeeny \cite{McWeeny}, we can reformulate the DFT by
ensuring not only that a variational procedure leads to
$\rho ({\bf r})$ ---which is derivable from a wave-function
$\Psi({\bf r}_1s_1,\ldots,{\bf r}_Ns_N)$ (the so-called
$N$-representability problem)---, but also the wave function belongs
to the totally irreducible representation of the spin permutation group
$A$.
From a mathematical point of view, the above proposition can be
expressed (in atomic units) as
\begin{eqnarray}
 E & = &
  \underset{\rho \to \gamma_1\ {\rm derived}\ {\rm from}\ \Psi \in A}{\rm min}
  \Bigg\{ - \frac{1}{2} \int_{{\bf r}_1 = {\bf r}_1'}
  \left[ \nabla \cdot \nabla^T \gamma_1({\bf r}_1;{\bf r}_1') \right]
   \! d{\bf r}_1
 \nonumber \\ & &
  + \int v_{\rm ext} ({\bf R},{\bf r}_1) \ \! \gamma_1 ({\bf r}_1)
   \ \! d{\bf r}_1
  + \frac{1}{2} \int \frac{\gamma_1({\bf r}_1)
   \ \! (\mathbb{I} - \hat{P}_{12})
   \ \! \gamma_1({\bf r}_2;{\bf r}_2')}{r_{12}}
    \ d{\bf r}_1 \ \! d{\bf r}_2
 \nonumber \\ & &
  + \underset{\gamma_2\ {\rm derived}\ {\rm from}\ \Psi \in A}{\rm min}\
  E_{\rm corr} [\gamma_2 ({\bf r}_1,{\bf r}_2)] \Bigg\} .
 \label{eq7tca}
\end{eqnarray}
This equation shows the relationship between the one-electron function,
$\gamma_1({\bf r}_1;{\bf r}_1')$, and the main part of the energy
functional ---the rest of the functional, which is the
electron-electron repulsion, depends on $\gamma_2({\bf r}_1,{\bf r}_2)$.
The last term is also a functional of the one-electron density.
In the new reformulation of DFT, the methodology is almost universally
based on the Kohn-Sham approach and only differs in the particular way
to model the unknown ``exchange-correlation'' term.


\section{Time-Dependent Density Functional Theory}
\label{sec3}


An extension of standard DFT is its TD version.
This generalization is necessary when dealing with intrinsic TD
phenomena.
In addition, it preserves the appealing flavor of the classical
approach to the theory of motion.

The rigorous foundation of the TD-DFT was started with the works by
Bartolotti \cite{Bartolotti} and Deb and Ghosh \cite{Deb}.
However, the proofs of the fundamental theorems were provided by Runge
and Gross \cite{Runge}.
One of those theorems corresponds to a Hohenberg-Kohn-like theorem for the TD Schr\"odinger equation.
The starting point for the derivation of the TD Kohn-Sham (KS)
equations is the variational principle for the quantum mechanical
action (along this Section, atomic units are also used):
\begin{equation}
 S[\Psi] = \int_{t_0}^{t_1}
  \langle \Psi (t) | \left[ i \ \! \frac{\partial}{\partial t}
   - \hat{H}(t) \right] | \Psi (t) \rangle \ \! dt .
 \label{xavi1}
\end{equation}
This variational principle is not based on the total energy
because in TD systems the total energy is not conserved.
The so-called Runge-Gross theorem then states that there exists a
one-to-one mapping between the external potential (in general, TD),
$v_{ext}({\bf r},t)$, and the electronic density, ${\rho}({\bf r},t)$,
for many-body systems evolving from a fixed initial state, $\Psi (t_0$.
Runge and Gross thus open the possibility of rigorously
deriving the TD version of the Kohn-Sham equations.
This procedure yields the TD Schr\"{o}dinger equation for the Kohn-Sham
electrons described by the orbitals $\phi_k({\bf r},t)$,
\begin{equation}
 i \ \! \frac{\partial \phi_k ({\bf r},t)} {\partial t} =
  H_{\rm KS}({\bf r},t) \ \! \phi_k ({\bf r},t) ,
 \label{xavi2}
\end{equation}
where the KS Hamiltonian is
\begin{equation}
 H_{\rm KS}({\bf r},t) = - \frac{1}{2} \ \! \nabla^2
  + v_{\rm KS}[{\rho}({\bf r},t)] ,
 \label{xavi3}
\end{equation}
with a TD-KS effective potential, usually given by the sum of three
terms, which account for external, classical electrostatic and
exchange interactions.
The latter is the source of all non-trivial, non-local, strongly
correlated many-body effects.

By construction, the exact TD density of the interacting system can
then be calculated from a set of non-interacting, single-particle
orbitals fulfilling the TD-KS equation (\ref{xavi2}) and reads
\begin{equation}
 \rho({\bf r},t)=\sum_{k=1}^N |\phi_k({\bf r},t)|^2 .
 \label{xavi4}
\end{equation}
Further analysis from the minimum action principle shows that the
exchange (xc) potential is then the functional derivative of that
quantity in terms of the density,
\begin{equation}
 v_{\rm xc} ({\bf r},t) = \frac{\delta S_{\rm xc}}
  {\delta \rho({\bf r},t)} ,
 \label{xavi5}
\end{equation}
where $S_{\rm xc}$ includes all non-trivial many-body parts of the
action.
The above equations provide the starting ground for further derivations
of the theory.
Thus, in addition to the TD-KS scheme, other variants have been
proposed across the years, which include the TD spin-DFT, the TD
current-DFT, the TD linear response DFT and the basis-set
DFT \cite{Marques}.
Each method has its range of applicability, but discussing them is out
of the scope of this chapter.

Here we focus on yet another implementation, the {\it single-particle}
hydrodynamic approach or QFD-DFT, which it provides a natural link
between DFT and Bohmian trajectories.
The corresponding derivation is based on the realization that the
density, ${\rho}({\bf r},t)$, and the current density,
${\bf j}({\bf r},t)$
%
%
%
satisfy a coupled-set of ``classical fluid'', Navier-Stokes equations:
\begin{eqnarray}
 \frac{\partial \rho ({\bf r},t)}{\partial t} & = &
 - \nabla {\bf j}({\bf r},t) ,
 \label{xavi7b} \\
 \frac{\partial {\bf j} ({\bf r},t)}{\partial t} & = &
  {\bf P}[\rho]({\bf r},t) ,
 \label{xavi7}
\end{eqnarray}
with
\begin{equation}
 {\bf P}[\rho]({\bf r},t) =
  - i \langle \Psi [\rho](t) | [{\bf j}({\bf r}),H(t)] |
\Psi [\rho](t) \rangle ,
 \label{xavi8}
\end{equation}
being a functional of the density and with initial conditions
$\rho ({\bf r},t_0)$ and ${\bf j} ({\bf r},t_0)$.

One can finally show that the above coupled equations translate
into one single particle nonlinear differential equation for the
hydrodynamical wave function
${\Phi}({\bf r},t)={\rho}({\bf r},t)^{1/2}
{\rm e}^{iS({\bf r},t)}$ in terms of potential energy functionals:
\begin{equation}
 \left (- \frac{1}{2} \nabla^2 + v_{\rm eff} [\rho] \right )
  \Phi ({\bf r},t)
  = i \ \! \frac{\partial \Phi ({\bf r},t)}{\partial t} ,
 \label{xavi9}
\end{equation}
with $v_{\rm eff}[{\rho}]$ being given by
\begin{equation}
 v_{\rm eff}[{\rho}] = \frac{\delta E_{\rm el-el}}{\delta \rho}
  + \frac{\delta E_{\rm nu-el}}{\delta \rho} +
 \frac{\delta E_{\rm xc}}{\delta \rho}+ \frac{\delta T_{\rm corr}}
 {\delta \rho} + \frac{\delta E_{\rm ext}}{\delta \rho}
 \label{xavi10}
\end{equation}
and where ${\bf j}({\bf r},t)={\rho}({\bf r},t)
{\bf v}({\bf r},t)$, with ${\nabla}S({\bf r},t)= {\bf v}({\bf r},t)$.
For many particle systems, this is still an open problem
(see next Section for a new discussion).
In Eq. (\ref{xavi10}), each term corresponds, respectively, to the
interelectronic repulsion energy, the Coulomb nuclear-electron
attraction energy, the exchange and correlation energy, the
nonclassical correction term to Weizs\"acker's kinetic energy,
and the electron-external field interaction energy functionals.
A judicious choice in the form of the above functionals yields
surprisingly good results for selected applications.

As a simple mathematical approach to QFD-DFT,
let us consider that the $N$-electron system is described by the TD
orbitals $\phi_k ({\bf r},t)$ when there is an external periodic, TD
potential, for which we want to obtain the (TD) density
$\rho({\bf r},t)$. These orbitals can be expressed in polar form,
\begin{equation}
 \phi_k ({\bf r},t) = R_k ({\bf r},t) \ \! {\rm e}^{iS_k ({\bf r},t)} ,
 \label{pra5}
\end{equation}
where the amplitudes $R_k ({\bf r},t)$ and phases $S_k ({\bf r},t)$
are real functions of space and time, and the former are subject to
the normalization condition
\begin{equation}
 \int_t \int R_k({\bf r},t) R_l({\bf r},t) d{\bf r} = \delta_{kl} ,
 \label{pra8}
\end{equation}
where $\int_t$ denotes the time-averaged integration over one period of
time.
The kinetic energy associated with this (noninteracting) $N$-electron
system reads now \cite{Bartolotti} as
\begin{equation}
 T_s [\{R_k,S_k\}]_t = - \frac{1}{2}
 \sum_{k=1}^N \int_t \int
 \left\{ R_k({\bf r},t) [\nabla^2 R_k({\bf r},t)]
 - R_k({\bf r},t) [\nabla S_k({\bf r},t)]^2 \right\} d{\bf r} .
 \label{pra6}
\end{equation}
Similarly to the time-independent case, here we also assume the
constraint that the sum of the squares of the $R_k$ gives the exact
density $\rho ({\bf r},t)$, i.e.,
\begin{equation}
\sum_{k=1}^N R_k^2 ({\bf r},t) = \rho ({\bf r},t) .
\label{pra7}
\end{equation}
Moreover, we introduce an additional constraint: the conservation of
the number of particles,
\begin{equation}
\sum_{k=1}^N \frac{\partial R_k^2}{\partial t}
\left( = \frac{\partial \rho}{\partial t} \right) =
- \nabla \cdot {\bf j} ,
\label{pra9}
\end{equation}
where ${\bf j}$ is the single-particle quantum density current vector.
After minimizing Eq.~(\ref{pra6}) with respect to the $R_k$ (which is
subject to the previous constraints), we reach the Euler-Lagrange
equation
\begin{equation}
 - \frac{1}{2} \nabla^2 R_k + v_{\rm eff} R_k = \epsilon_k R_k ,
\label{pra10}
\end{equation}
where $v_{\rm eff} ({\bf r},t)$ and $\epsilon_k({\bf r},t)$ are
the Lagrange multiplier associated with the constraint defined in
Eq.~(\ref{pra7}) and the conservation of the number of particles
given by Eqs.~(\ref{pra8}) and (\ref{pra9}), respectively.
Moreover, $\epsilon_k ({\bf r},t)$ can be split up as a sum of two
terms
\begin{equation}
 \epsilon_k ({\bf r},t) = \epsilon_k^{(0)}
  + \epsilon_k^{(1)} ({\bf r},t) .
 \label{pra11}
\end{equation}
The quantity $\epsilon_k^{(0)}$ is a result of the normalization
constraint, while $\epsilon_k^{(1)}$ are the Lagrange multipliers
associated with the charge-current conservation defined by
Eq.~(\ref{pra9}). On the other hand, if Eq.~(\ref{pra10}) is
divided by $R_k$ we can reexpress the corresponding equation as
\begin{equation}
 Q_k ({\bf r},t) + v_{\rm eff} ({\bf r},t)= \epsilon_k ({\bf r},t)
 \label{pra}
\end{equation}
where $Q_k$ is the so-called quantum potential associated with the
state $\phi_k$,
\begin{equation}
 Q_k ({\bf r},t) = - \frac{1}{2} \frac{\nabla^2 R_k}{R_k} .
 \label{pra10bbis}
\end{equation}

Next, we minimize $T_s [\{R_k,S_k\}]_t$ with respect to $S_k$ to
be subject to the constraint
\begin{equation}
 \frac{\partial S_k}{\partial t} = - \epsilon_k ({\bf r},t) .
 \label{pra12}
\end{equation}
The resulting Euler-Lagrange equation is given by
\begin{equation}
 \frac{\partial R_k^2}{\partial t}
  + \nabla \cdot (R_k^2 \, \nabla S_k) = 0 .
 \label{pra13}
\end{equation}
The coupled equations, Eqs.~(\ref{pra10}) and (\ref{pra13}), provide
a means of determining the exact TD density of the system of interest.
We note that, at the solution point, the current vector is given by
\begin{equation}
{\bf j} ({\bf r},t) =
\sum_{k=1}^N R_k^2 ({\bf r},t) \nabla S_k ({\bf r},t) .
\label{nopra}
\end{equation}

Note that, in the limit that the time-dependence is turned off, the
TD-DFT approach correctly reduces to the usual time-independent DFT
one, since $\nabla S_k$ vanishes, Eqs.~(\ref{pra9}), (\ref{pra12}) and
(\ref{pra13}) are identically satisfied, and Eq.~(\ref{pra6}) will
reduce to the time-independent kinetic energy of an $N$-electron
system.


\section{Bohmian mechanics. A trajectory picture of quantum mechanics}
\label{sec4}


\subsection{Single-particle trajectories}

Apart from the operational, wave or action-based pictures of Quantum
Mechanics provided by Heisenberg, Schr\"odinger or Feynman,
respectively, there is an additional, fully trajectory-based picture:
Bohmian mechanics \cite{Bohm,Holland}.
Within this picture, the standard quantum formalism is understood in
terms of trajectories defined by very specific motion rules.
Although this formulation was independently formulated by Bohm, it
gathers two former conceptual ideas: (1) the QFD picture proposed by
Madelung, and (2) the pilot role assigned to the wave function,
proposed by de Broglie.
In this way, the time-evolution or dynamics of the system is described
as an ideal quantum fluid with no viscosity; the evolution of this flow
of identical particles is ``guided'' by the wave function.

The Bohmian formalism follows straightforwardly from the Schr\"odinger
one in the position representation after considering a change of
variables, from the complex wave function field ($\Psi,\Psi^*$) to
the real fields ($\rho,S$) according to the transformation relation:
\begin{equation}
\Psi ({\bf r},t) =
R({\bf r},t) \ \! e^{i S({\bf r},t)/\hbar} ,
\label{eq:2}
\end{equation}
with $\rho = R^2$.
Substituting this relation into the TD Schr\"odinger equation for a
single particle of mass $m$,
\begin{equation}
i \hbar \ \! \frac{\partial \Psi({\bf r},t)}{\partial t}
= \left[ -\frac{\hbar^2}{2m} \ \nabla^2 +
V({\bf r}) \right] \Psi({\bf r},t) ,
\label{eq:1}
\end{equation}
and then separating the real and imaginary parts from the resulting
expression, two real coupled equations are obtained:
\begin{subequations}
\begin{eqnarray}
\frac{\partial \rho}{\partial t} & + &
\nabla \! \cdot \!
 \left( \rho \ \frac{\nabla S}{m} \right) = 0 ,
\label{eq:3a} \\
\frac{\partial S}{\partial t} & + &
\frac{(\nabla S)^2}{2m} + V_{\rm eff} = 0 ,
\label{eq:3b}
\end{eqnarray}
\label{eq:3}
\end{subequations}
where
\begin{equation}
V_{\rm eff} = V + Q = V - \frac{\hbar^2}{2m}
\frac{\nabla^2 R}{R}
= V - \frac{\hbar^2}{4m} \left[
\frac{\nabla^2 \rho}{\rho} - \frac{1}{2}
\left( \frac{\nabla \rho}{\rho} \right)^2 \right]
\label{eq:4}
\end{equation}
is an \emph{effective potential} resulting from the sum of the
``classical'' contribution, $V$, and the so-called \emph{quantum
potential}, $Q$, which depends on the quantum state via $\rho$
---or, equivalently, on the instantaneous curvature of the wave
function via $R$.
Note that in the case $V = v_{\rm eff}$ and $\Psi$ given as in the
previous Section, one gets $V_{\rm eff} = \epsilon_k$ or $V_{\rm eff}
= \epsilon_k ({\bf r},t)$ depending on whether we are considering the
time-independent or the TD case, respectively.
The action of the whole ensemble through the wave function
on the particle motion can be seen as  a dynamical
manifestation of quantum nonlocality.
Equation (\ref{eq:3a}) is the continuity equation for the particle
flow (or the probability density, from a conventional viewpoint)
and (\ref{eq:3b}) is a \emph{generalized} (quantum) Hamilton-Jacobi
equation.
As in classical mechanics, the \emph{characteristics} or solutions,
$S$, of Eq.~(\ref{eq:3b}) define the particle velocity field,
\begin{equation}
 {\bf v} = \frac{\nabla S}{m} ,
 \label{eq:5}
\end{equation}
from which the \emph{quantum trajectories} are known.
\emph{Uncertainty} arises from the \emph{unpredictability} in
determining the particle initial conditions ---distributed according
to $\rho ({\bf r}, t=0)$ \cite{Holland}---, but not from the
\emph{impossibility} to know the actual (quantum) trajectory
pursued during its evolution.

An alternative way to obtain the quantum trajectories is by formulating
Bohmian mechanics as a Newtonian-like theory.
Then, Eq.~(\ref{eq:5}) gives rise to a \emph{generalized} Newton's
second law,
\begin{equation}
 m \ \! \frac{d {\bf v}}{dt} = - \nabla V_{\rm eff} .
 \label{eq:6}
\end{equation}
This formulation results very insightful; according to
Eq.~(\ref{eq:6}), particles move under the action of an
\emph{effective force}, $- \nabla V_{\rm eff}$, i.e., the nonlocal
action of the quantum potential here is seen as the effect of a
(nonlocal) quantum force.
From a computational viewpoint, this formulation results very
interesting in connection to quantum hydrodynamics \cite{Madelung,Wyatt}. Thus, Eqs.~(\ref{eq:3}) can be reexpressed in
terms of a continuity equation and a  \emph{generalized} Euler
equation.
%
%
As happens with classical fluids, here also two important concepts come
into play: the \emph{quantum pressure} and the {\it quantum vortices}
\cite{vortices} which occur at nodal regions where the velocity field
is rotational.

Since TD-DFT is being applied to scattering problems in its QFD
version, two important consequences of the nonlocal nature of the
quantum potential worth stressing in this regard.
First, relevant quantum effects can be observed in regions where the
classical interaction potential $V$ becomes negligible and, more
important, where $\rho({\bf r},t) \approx 0$.
This happens because quantum particles respond to the ``shape'' of
$\Psi$, but not to its ``intensity'', $\rho ({\bf r},t)$ ---notice that
$Q$ is scale-invariant under the multiplication of $\rho ({\bf r},t)$
by a real constant.
Second, quantum-mechanically the concept of \emph{asymptotic} or
\emph{free motion} only holds \emph{locally}.
Following the classical definition for this motional regime,
\begin{equation}
m \ \! \frac{d {\bf v}}{dt} \approx 0 ,
\label{eq:7}
\end{equation}
this means in Bohmian mechanics that $\nabla V_{\rm eff} \approx 0$,
i.e., the local curvature of the wave function has to be zero (apart
from the classical-like requirement that $V \approx 0$).
In scattering experiments this condition is satisfied along the
directions specified by the diffraction channels \cite{channels};
in between, although $V \approx 0$, particles are still subject
to strong quantum forces.


\subsection{Bohmian trajectories describing many-body systems}
\label{sec4b}

In the case of a many-body problem, the Bohmian mechanics for an
$N$-body dynamics follows straightforwardly from the one for a single
system, but replacing Eq.~(\ref{eq:2}) by
\begin{equation}
\Psi ({\bf r}_1,{\bf r}_2,\ldots,{\bf r}_N;t) =
 R({\bf r}_1,{\bf r}_2,\ldots,{\bf r}_N;t)
\ \! e^{i S({\bf r}_1,{\bf r}_2,\ldots,{\bf r}_N;t)/\hbar} ,
\label{eq:2bis}
\end{equation}
with $\rho ({\bf r}_1,{\bf r}_2,\ldots,{\bf r}_N;t) =
R^2 ({\bf r}_1,{\bf r}_2,\ldots,{\bf r}_N;t)$.
If we are interested in the density of a single particle, we need
to ``trace'' over the remaining $N$$-$1 degrees of freedom in the
corresponding density matrix (see Sec.~\ref{sec4c}).
On the other hand, in order to know the specific trajectory pursued by
the particle associated with the $k$th degree of freedom, we have to
integrate the equation of motion
\begin{equation}
{\bf v}_k = \frac{\nabla_k S}{m} ,
\label{eq:5bis}
\end{equation}
where $\nabla_k = \partial/\partial {\bf r}_k$. The velocity field
is irrotational in nature except at nodal regions.
Obviously, there will be as many equations of motion as degrees of
freedom.
Note that since each degree of freedom represents a particle that
is interacting with the remaining $N$$-$1 particles in the ensemble, the
corresponding trajectory will be strongly influenced by the evolution
of those other $N$$-$1 particles.
This {\it entanglement} is patent through the quantum potential, which
is given here as
\begin{equation}
Q = - \frac{\hbar^2}{2m} \sum_{k=1}^N \frac{\nabla_k^2 R}{R} ,
\label{qpot1}
\end{equation}
where $Q = Q({\bf r}_1,{\bf r}_2,\ldots,{\bf r}_N;t)$ is nonseparable
and, therefore, strongly nonlocal.
Note that this nonlocality arises from correlation among different
degrees of freedom, which is different from the nonlocality that
appears when considering symmetry properties of the wave function,
not described by the Schr\"odinger equation but by quantum
statistics. In this sense, we can speak about two types of
entanglement: symmetry and dynamics. The general $N$-body wave
function ({\ref{eq:2bis}) is entangled in both aspects.

Now, if the many-body (electron) problem
can be arranged in such a way that the many-body, nonseparable wave
function is expressed in terms of a separable wave function which
depends on $N$ single-particle wave functions
(Hartree approximation), i.e.,
\begin{equation}
\Psi ({\bf r}_1,{\bf r}_2,\ldots,{\bf r}_N;t) =
\Pi_{k=1}^N \psi_k ({\bf r}_k;t) =
\Pi_{k=1}^N R_k ({\bf r}_k;t) \ \! e^{i S_k({\bf r}_k;t)/\hbar} ,
\label{eq:2bis2}
\end{equation}
%
then, in terms of trajectories, we find a set of uncoupled equations
of motion,
\begin{equation}
{\bar {\bf v}}_k = \frac{\nabla_k S_k}{m} ,
\label{eq:5bis2}
\end{equation}
which will only depend implicitly (through $v_{\rm eff}$) on the other
particles. Note that the {\it factorization} of the wave function
implies that the quantum potential becomes a separable function
of the $N$ particle coordinates and time,
\begin{equation}
Q = - \frac{\hbar^2}{2m} \sum_{k=1}^N \frac{\nabla_k^2 R_k}{R_k}
= \sum_{k=1}^N Q_k ,
\label{qpot12}
\end{equation}
where each $Q_k$ measures the local curvature of the wave function
associated with the $i$th orbital associated to the corresponding
particle.
Therefore, each degree of freedom can be studied separately from the
rest (with the exception that we have to take into account the mean
field created by the remaining $N$$-$1 particles).
Factorizability implies physical independence, statistical
independence or, in other words, that particles obey Maxwell-Boltzmann
statistics (they are distinguishable) and the associate wave
function is, therefore, not entangled.

In TD-DFT, the wave function is antisymmetrized and, therefore,
nonfactorizable or entangled. However, as said above, it is not
entangled from a dynamical point of view because the quantum
forces originated from a nonseparable  quantum potential like
Eq.~(\ref{qpot1}) are not taken into account .



\subsection{The reduced quantum trajectory approach}
\label{sec4c}

In Sec.~\ref{sec4b}, we have considered the problem of the reduced
dynamics from a standard DFT approach, i.e., in terms of
single-particle wave functions from which the (single-particle)
probability density is obtained.
However, one could also use an alternative description which arises
from the field of decoherence.
Here, in order to extract useful information about the system of
interest, one usually computes its associated reduced density matrix
by tracing the total density matrix, $\hat{\rho}_t$ (the subscript $t$
here indicates time-dependence), over the environment degrees of
freedom.
In the configuration representation and for an environment constituted
by $N$ particles, the system reduced density matrix is obtained after
integrating $\hat{\rho}_t \equiv |\Psi\rangle_t\ \! _t\langle\Psi|$
over the 3$N$ environment degrees of freedom, $\{{\bf r}_k\}_{k=1}^N$,
\begin{equation}
 \tilde{\rho} ({\bf r},{\bf r}';t) =
  \int \langle {\bf r},{\bf r}_1, {\bf r}_2, \ldots {\bf r}_N |
  \Psi (t)\rangle
 \langle \Psi (t)| {\bf r}', {\bf r}_1, {\bf r}_2,
  \ldots {\bf r}_N \rangle \ \! {\rm d}{\bf r}_1 {\rm d}{\bf r}_2
  \cdots {\rm d}{\bf r}_N .
 \label{eq7}
\end{equation}
The system (reduced) quantum density current can be derived from this
expression, being
\begin{equation}
 \tilde{\bf j}({\bf r},t) \equiv \frac{\hbar}{m}
  \ {\rm Im} [ \nabla_{\bf r} \tilde{\rho} ({\bf r},{\bf r}';t)]
  \Big\arrowvert_{{\bf r}' = {\bf r}} ,
 \label{eq12}
\end{equation}
which satisfies the continuity equation
\begin{equation}
 \dot{\tilde{\rho}} + \nabla \tilde{\bf j} = 0 .
 \label{eq13}
\end{equation}
In Eq.~(\ref{eq13}), $\tilde{\rho}$ is the diagonal element (i.e.,
$\tilde{\rho} \equiv \tilde{\rho} ({\bf r},{\bf r};t)$) of the reduced
density matrix.
Taking into account Eqs.~(\ref{eq12}) and (\ref{eq13}), now we define
the velocity field, $\dot{\bf r}$, associated with the (reduced) system
dynamics as
\begin{equation}
 \tilde{\bf j} = \tilde{\rho} \dot{\bf r} ,
 \label{vfield}
\end{equation}
which is analogous to the Bohmian velocity field.
Now, from Eq.~(\ref{vfield}), we define a new class of quantum
trajectories as the solutions to the equation of motion
\begin{equation}
\dot{\bf r} \equiv \frac{\hbar}{m}
\frac{{\rm Im} [ \nabla_{\bf r} \tilde{\rho} ({\bf r},{\bf r}';t)]}
 {{\rm Re} [ \tilde{\rho} ({\bf r},{\bf r}';t)]}
 \Bigg\arrowvert_{{\bf r}' = {\bf r}} .
\label{eq14}
\end{equation}
These new trajectories are the so-called {\it reduced quantum
trajectories} \cite{SB}, which are only explicitly related to the
system reduced density matrix.
The dynamics described by Eq.~(\ref{eq14}) leads to the correct
intensity (whose time-evolution is described by Eq.~(\ref{eq13}))
when the statistics of a large number of particles is considered.
Moreover, it is also straightforward to show that Eq.~(\ref{eq14})
reduces to the well-known expression for the velocity field in Bohmian
mechanics when there is no interaction with the environment.


\section{Final discussion and conclusions}
\label{sec5}

Nowadays the success of DFT and TD-DFT is out of question in both the
Physics and Chemistry communities.
The numerical results obtained are most of cases in good agreement to
those issued from experimental and other theoretical methods with a
relative small computational effort.
However, in this chapter, our goal has been to present the TD-DFT from
a Bohmian perspective and to analyze, from a conceptual level, some of
the aspects which are deeply rooted in DFT.

Working with a system of fermions, where the total wave
function has to be antisymmetrized with respect to
two-particle interchanges, it gives rise to the appearance of new quantum
forces from the quantum potential which are not described by the DFT
Hamiltonian.
The DFT wavefunction will be then nonfactorizable and, therefore,
entangled from a symmetry point of view but not from a dynamical
point of view.
In this sense, as mentioned above, the effective potential
$V_{\rm eff}$ plays a fundamental role not only in the nonlocality
of the theory, but in the so-called invertibility problem of the
one-to-one mapping up to an additive time-dependent function between
the density and $v_{\rm eff}$.
In our opinion, the central theorems of TD-DFT should be written
in terms of $V_{\rm eff}$ instead of $v_{\rm eff}$, since the
quantum potential is also
state-dependent and a functional of the density.
An infinite set of possible quantum potentials can be associated with
the same physical situation and Schr\"odinger equation, and therefore
the invertibility should be questioned.
Moreover, for scattering problems, when $v_{\rm eff}$ is
negligible in the asymptotic region, the quantum potential can be
still active and the time propagation should be extended much
farther in order to obtain a good numerical convergence.

In Bohmian mechanics, the way how the full problem is tackled in order
to obtain operational formulas can determine dramatically the final
solution due to the context-dependence of this theory.
More specifically, developing a Bohmian description within the
many-body framework and then focusing on a particle is not equivalent
to directly starting from the reduced density matrix or from the
one-particle TD-DFT equation.
Being well aware of the severe computational problems coming from the
first and second approaches, we are still tempting to claim that those
are the most natural ways to deal with a many-body problem in a Bohmian context.


\section*{Acknowledgements}

This work was supported by the Ministerio de Ciencia e Innovaci\'on
(Spain) under Projects FIS2007-62006 and CTQ2005-01117/BQU and
Generalitat de Catalunya under Projects 2005SGR-00111 and
2005SGR-00175.
A.\ S.\ Sanz acknowledges the Consejo Superior de Investigaciones
Cient\'{\i}ficas for a ``Juan de la Cierva'' Contract.


\end{document}